\documentclass[conference]{IEEEtran}
\IEEEoverridecommandlockouts
\usepackage{cite}
\usepackage{amsmath,amsfonts,amssymb,amsthm}
\usepackage{algorithmic}
\usepackage{graphicx}
\usepackage{textcomp}
\usepackage{xcolor}
\usepackage{enumitem}
\newtheorem{definition}{Definition}[section]
\theoremstyle{definition}
\usepackage{booktabs}
\usepackage{svg}
\usepackage{array}

\def\BibTeX{{\rm B\kern-.05em{\sc i\kern-.025em b}\kern-.08em
    T\kern-.1667em\lower.7ex\hbox{E}\kern-.125emX}}
\begin{document}

\title{Privacy Preserving Machine Learning Workflow: from Anonymization to Personalized Differential Privacy Budgets in Federated Learning
\thanks{\textit{Accepted at the 2nd International Conference on Federated Learning and Intelligent Computing Systems (FLICS2026).}}}

\author{\IEEEauthorblockN{1\textsuperscript{st} Judith Sáinz-Pardo Díaz}
\IEEEauthorblockA{\textit{Instituto de Física de Cantabria}\\
\textit{(IFCA) CSIC-UC}\\
Santander, Spain \\
sainzpardo@ifca.es\\
0000-0002-8387-578X}
\and
\IEEEauthorblockN{2\textsuperscript{nd} Álvaro López García}
\IEEEauthorblockA{\textit{Instituto de Física de Cantabria}\\
\textit{(IFCA) CSIC-UC}\\
Santander, Spain \\
aloga@ifca.es\\
0000-0002-0013-4602}}

\maketitle

\begin{abstract}
The growing development of artificial intelligence based solutions, together with privacy legislation, has driven the rise of the so-called privacy preserving machine learning architectures, such as federated learning. While federated learning enables model training on decentralized data preventing their sharing and centralization, it still faces several challenges related to data integrity and privacy. This paper presents a comprehensive privacy preserving federated learning workflow for sensitive tabular data, including anonymization and differential privacy techniques. We also introduce a formal definition for the concept of client drift, together with ways of detecting it to mitigate poisoning attacks. Then, we detail a complete methodology for assigning personalized privacy budgets for global differential privacy to the different clients participating in the network, based on a re-identification risk metric. The proposed methodology is presented and tested on an openly available dataset of medical records. Within the experimental setup we show that the approach based on personalized budgets, compared to the architecture including global differential privacy with fixed privacy budget, achieves a better model performance in terms of two error metrics. 
\end{abstract}


\section{Introduction}
It is widely known that artificial intelligence (AI) methods, especially machine and deep learning (ML/DL), have a wide range of useful applications \cite{8697857} in different fields, such as healthcare \cite{miotto2018deep}, where we deal with high sensitive data. Classically, in order to perform inference processes, centralized models are trained on as much data as possible, as a larger amount of data will allow the models to extrapolate better in the presence of unseen data. However, in many cases data that have been generated or collected by a device or an institution cannot be centralized due to privacy, security, or simply technical requirements. Concerning security and privacy issues, the General Data Protection Regulation (GDPR) \cite{GDPR2016a} comes into play. Although these types of techniques (AI, ML) are not explicitly mentioned in the GDPR, their scope of application has a direct impact on them in relation to the processing of personal data. In Article 5 of the GDPR, it is expressed within the principle of integrity and confidentiality, that data must be processed securely, including adequate protection against unauthorized or unlawful processing. In addition, the principles of data minimization (keeping no more information than the minimum amount of personal data necessary to fulfill the purpose for which it was collected), lawfulness, fairness and transparency, and storage limitation are also stated \cite{GDPR2016a}. 

The federated learning (FL) architecture emerges as a natural alternative to training machine or deep learning models in a centralized way, through distributed training on different clients \cite{mcmahan2017communication}. Going back to the specific requirements of the GDPR, a federated learning approach allows users to train models without the need to share or centralize data from multiple sources, which is in line with the principles of data integrity and confidentiality. However, while this type of distributed system does not compromise the security of the data in terms of its centralization and processing by an external or third party (as with the classical centralized approach) other privacy issues must be addressed. This involves the inclusion of additional privacy preserving techniques to ensure not only compliance with the GDPR but also that a potential attacker can extract the minimum possible information about the training data.

In this paper we propose a complete workflow for a process of privacy preserving machine learning training when dealing with sensitive tabular data. We review different privacy enhancing technologies (PETs) and their incorporation in FL architectures, allowing collaboration of different clients while incorporating additional privacy measures.

Different attacks on FL architectures and ways to mitigate them are reviewed and then a complete methodology is detailed. Within this methodology we propose a way to aggregate the weights by previously using a personalized private budget for differential privacy, based on a re-identification risk metric on the anonymized data. In addition, the concept of client drift and ways to detect it is formally defined and discussed. Finally, an experimental setup is carried out based on a use case of medical records, for which 10 clients are simulated based on the patients' country (thus emulating different hospital networks per country). Note that in this study the objective is to present the methodology in an experimentally reproducible use case. Thus, based on this work, researchers will be able to apply the proposed methodology to real-world applications, incorporating privacy accounting methodologies as well as stricter privacy levels.

Thus, this work is structured as follows: in Section~\ref{sec:fl} we introduce the basis of federated learning and different attacks. Then, Sections~\ref{sec:anonymity} and \ref{sec:dp} introduce anonymization methods for tabular data and differential privacy mechanism respectively, focusing on their incorporation in FL architectures. In Section~\ref{sec:drift} we formally define the concept of client drift and ways for detecting it on FL architectures. Section~\ref{sec:methodology} introduces the methodology proposed in this work for performing a complete privacy preserving machine learning workflow based on FL for sensitive tabular data. In this section we introduce the idea of applying personalized privacy budgets for differential privacy based on a re-identification risk metric. Then, in order to show the applicability of the workflow proposed, Section~\ref{sec:experimental} presents a synthetic experimental setup based on openly available data. Finally, Section~\ref{sec:future} draws the conclusions and future work.

\section{Federated Learning}\label{sec:fl}
The goal of a federated learning (FL) architecture is twofold: (1) to train the models with as much and as diverse data as possible; (2) to do it in a distributed way without having to centralize or share the data with different parties (preserving privacy). 

In this sense, the architecture \cite{ZHANG2021106775} consists of a central server and several clients (data owners) following the next scheme: (1) the clients agree on the model to be used, or it is created and given by the server; (2) each client receives the model to be trained; (3) each client trains the model locally; (4) the weights obtained after training locally are sent to the server; (5) the server aggregates all the weights received; (6) the model is updated with the aggregated weights  $\longrightarrow$ repeat from (2). 

Regarding step number (5), the aggregation of the models can be performed using different strategies. The most widely used is \textit{FedAvg}, which will be adopted in this study and that is defined as the aggregated average according to the number of training data of each client. 

The use cases that take advantage of the adoption of a FL architecture are quite interdisciplinary, but in general they are motivated by the need to preserve the privacy of data from different data providers by not centralizing them or sharing them with third parties. In this sense, numerous applications in the medical field stand out \cite{guan2024federated, sohan2023systematic}, where it is required to work with sensitive data from different sources without data sharing or centralization.

Intuitively, it is evident that the privacy of the clients (data owners) participating in an FL architecture is preserved from the point of view that their data cannot be shared or centralized for training purposes. However, even if the data is not shared, the models developed are susceptible to attacks, also in terms of the privacy and integrity of the data. This requires the incorporation of PETs to add an additional security layer. 

Specifically, some of the most commonly used PETs in data science contexts involve anonymization and differential privacy. However, the application of these methodologies on a federated architecture presents substantial differences with respect to their integration in a classical centralized architecture (as will be detailed in the following sections in order to introduce the proposed methodology). Before moving on, Table~\ref{tab:attacks_fl} reviews some of the most common attacks on a federated learning architectures together with the main issue associated (see \cite{9945997, RODRIGUEZBARROSO2023148}). 

\begin{table}[ht]
    \centering
    \caption{Common attacks in federated learning architectures and main issue.}
    \label{tab:attacks_fl}
    \begin{tabular}{p{1.5cm}p{6.5cm}}
    \toprule
        \textbf{Attack} & \textbf{Main issue} \\
        \midrule
        \textit{Reconstruction} & An adversary can reconstruct the original private data by intercepting the model weights/gradients.\\
        \toprule
        \textit{Model inference} & The attacker extracts the sensitive information from the model trained such as the labels or specific attributes.\\
        \toprule
        \textit{Membership inference} & An attacker can determine if a particular data point belongs to certain client's training dataset.\\
        \toprule
        \textit{Poisoning} & A client damages the overall accuracy of the model intentionally (e.g. by altering the resulting model) or unintentionally (by errors in the data used in training, such as in labeling).\\
    \bottomrule
    \end{tabular}
\end{table}

\section{Data anonymization in federated learning}\label{sec:anonymity}
The general data protection regulation states that data that has been appropriately anonymized is no longer considered personal data about an identified or identifiable person \cite{GDPR2016a}. This is because data anonymization techniques are usually applied on tabular data to make an individual in a database unidentifiable, for which it is essential to first remove those identifying attributes, and apply hierarchical transformations to the remaining quasi-identifiers\footnote{The quasi-identifiers are the attributes that by themselves do not allow the identification of an individual, but in combination with each other they can make it possible to identify an individual.}. 

Some of the most common anonymity techniques include \textit{k-anonymity}, \textit{$\ell$-diversity}, \textit{t-closeness}, \textit{$\beta$-likeness}, \textit{$\delta$-disclosure}, etc. These methods are applied on tabular data in which we can locate quasi-identifiers and sensitive attributes. For example, given an equivalence class (EC) consisting of all the rows of the dataset that are indistinguishable from the set of quasi-identifiers, \textit{$k$-anonymity} is verified if all these equivalence classes have at least $k$ rows, and \textit{$\ell$-diversity} if each EC has $\ell$ different values for the sensitive attribute (in the case there is only one sensitive attribute, see \cite{sainz2022python} for the case of more than one). In addition, an equivalence class verifies \textit{t-closeness} \cite{domingo2016database} if the distribution of the values of the sensitive attribute is at a distance no closer than $t$ from its distribution in the whole database. Then, given a sensitive attribute, a database verifies \textit{t-closeness} if all the equivalence classes verify it.

To apply these techniques successfully, it is necessary to carry out different transformations on the data, in particular on the values in the different columns (usually only on the quasi-identifiers). For this purpose, different hierarchy trees are established that enable us to generalize over the initial values. In particular, given such a hierarchy tree, we will obtain a latent space of solutions according to the applied transformations. 

In federated learning schemes, clients may apply different transformations to their data in order to verify certain privacy constraints that prevent knowledge extraction, which may occur even from the weights that define a model. It is important that in order to train the models and subsequently aggregate them, the input data in all clients are generalized to the same levels. 

Let $N$ be the number of clients, $m$ the number of features, and $h_{ij}$ the hierarchy level applied to feature $j$ of client $i$. Each client $i$ will send to the server before starting the federated learning process a vector as follows: $(h_{i1}, h_{i2}, \hdots, h_{im})$. Once the server receives such information from all the clients it will take the hierarchy level that allows that in all the cases the privacy constraint imposed by all the clients are (at least) fulfilled. Then, the server will send back to each client a vector $\overline{h}=(\overline{h_{1}}, \overline{h_{2}}, \hdots, \overline{h_{m}})$, with $\overline{h_{j}}=max_{i\in\{1,\hdots, N\}}h_{ij}$ $\forall j \in \{1,\hdots,m\}$. Each client must apply the hierarchy level given in the vector $\overline{h}$ before proceeding to start the federated training.

In this context, the \textit{anjana} Python library \cite{sainz2024open} can be used to first apply the desired anonymization techniques locally for each database. Then, given the transformations performed to meet each client’s privacy levels, $\overline{h}$ can be calculated and applied to their data. Thus, all datasets are generalized to the same level within the latent space and ready for a unified federated learning architecture.

Coming back to Table~\ref{tab:attacks_fl}, let us note that, intuitively, by applying a proper anonymization of the data we will be reducing the risk model inference attacks in the sense that an attacker will not be able to extract sensitive information from the training data because, in case he/she can retrieve it, it will have been properly anonymized to prevent such disclosure according to the established privacy thresholds. However, models applied to anonymized data remain vulnerable, making it necessary to include additional protection measures such as differential privacy.

\section{Differentially private federated learning}\label{sec:dp}

As already mentioned, with anonymization we are protecting the raw data to avoid linkage attacks, re-identification or skewness among others (see \cite{sainz2022python}). Specifically, we focus directly on protecting the input of the model to be developed (the data). In this sense, \textit{input privacy} is a guarantee that one or more users can participate in a computational process so that none of the parties know anything about each other's inputs. Then, with input privacy we are just ensuring privacy regarding the input of an information flow. 

Furthermore, \textit{output privacy} tries to guarantee that the output of an information flow cannot be reversed to learn specific attributes about the input \cite{wang2011output}. An example of a tool for ensuring output privacy is \emph{differential privacy} (DP) \cite{dwork2014algorithmic}.

The identity of the users associated with databases that appear to be anonymized can be revealed in some cases by linkage attacks. The first approach to solve this problem via masking or alteration of the data is to introduce statistically controlled noise into the data. The more noise that is introduced, the more difficult it will be to break the privacy. However, it must be noted that adding too much noise can make the data unusable.  Since adding noise can considerably reduce the usefulness of the data, the natural idea when talking about differential privacy is to add noise statistically to the response of requests or the queries made on a database.

We define \textit{$\epsilon$-differential privacy} and \textit{($\epsilon$,$\delta)$-differential privacy} \cite{kim2021privacy} \cite{zhu2017differential} as follows:

\begin{definition}
A randomized algorithm $\mathcal{A}$, with domain $\mathcal{D}$ and range $\mathcal{R}$, satisfies \textit{$\epsilon$-differential privacy} if for any two adjacent inputs $X, X' \in \mathcal{D}$ and for any subset of outputs $\mathcal{Z}\subseteq \mathcal{R}$ it is satisfied that:

$$
\mathbb{P}[\mathcal{A}(X)\in \mathcal{Z}] \leq e^{\epsilon}\mathbb{P}[\mathcal{A}(X')\in \mathcal{Z}], \mbox{  with  } \epsilon\geq 0.
$$

In the same conditions, it satisfies \textit{$(\epsilon, \delta)${-differential privacy}} with $\epsilon\geq 0$ and $\delta \in [0,1]$ if:

$$
\mathbb{P}[\mathcal{A}(X)\in \mathcal{Z}] \leq e^{\epsilon}\mathbb{P}[\mathcal{A}(X')\in \mathcal{Z}] + \delta.
$$
\end{definition}

Some of the most common mechanisms to achieve verification of such definitions are the Laplace or Gaussian mechanisms. The definitions of such mechanisms for \textit{$\epsilon$-differential privacy} and \textit{($\epsilon$,$\delta$)-differential privacy} respectively are presented below. First, let us define the concepts of \textit{$l_{1}$-sensitivity} and \textit{$l_{2}$-sensitivity} \cite{zhu2017differential}:

\begin{definition}
Be $f:\mathcal{D} \longrightarrow \mathbb{R}^{k}$, we define the \textit{$l_{i}$-sensitivity} ($\Delta_{i}(f)$) for $i=1,2$ as follows:

$$
\displaystyle
\Delta_{i}(f):= \max_{||x-y||_1}||f(x)-f(y)||_{i}, \forall i \in \{1,2\}.
$$
\end{definition}

\begin{definition}
Given any function $f:\mathcal{D} \longrightarrow \mathbb{R}^{k}$, we define the \textit{Laplace Mechanism} as:
$$
\displaystyle
\mathcal{A}_{L}(x, f(\cdot), \epsilon):= f(x) + (Y_{1}, \hdots, Y_{k}),
$$
where $Y_{i},  \forall i \in \{1, \hdots, k\}$ are independent and identically distributed variables drawn from the distribution $Laplace(0, \Delta_{1} (f)/\epsilon)$.
\label{def:laplace}
\end{definition}

\begin{definition}
Given any function $f:\mathcal{D} \longrightarrow \mathbb{R}^{k}$, we define the \textit{Gaussian Mechanism} as:
$$
\displaystyle
\mathcal{A}_{G}(x, f(\cdot), \epsilon, \delta):= f(x) + (Y_{1}, \hdots, Y_{k}),
$$
where $Y_{i},  \forall i \in \{1, \hdots, k\}$ are independent and identically distributed (i.i.d.) variables from the Gaussian distribution $N(0, \sigma^2)$, with $\sigma = \frac{\Delta_{2}(f) \sqrt{2\log(1.25/\delta)}}{\epsilon}$.
\label{def:gaussian}
\end{definition}

In addition, we can distinguish between local differential privacy and global differential privacy. In the first case, the noise is added to the raw data, before it is centralized or shared with a server and therefore before performing queries on it, applying ML/DL models, or simply extracting statistics. On the other hand, in the case of global DP such noise is added to the queries that are performed on the data or during the procedures that are performed on them, such as during the training of a ML/DL model, for example during the gradient descent step in the case of an ANN \cite{zhu2017differential}. 

The need to include differential privacy in a federated learning architecture can be motivated by two scenarios: (1) the reliability of all the clients involved in the training is not guaranteed; (2) the FL server in charge of the aggregation is not trusted. We may even be in the case where (1) and (2) occur. 

In this work we will focus on the first case, assuming that the central server in charge of performing the aggregation in a FL architecture is trusted. In this scenario, global DP can be applied when aggregating the weights of the models with the selected metric or aggregation strategy.

Within a FL architecture we can add Gaussian noise during the aggregation process in order to prevent information extraction from the global aggregated model (in case of dealing with untrusted clients that may act as attackers). Then, be the clipping norm $C$, the noise multiplier ($n_{\epsilon}$) and the number of sampled clients $m$. We can apply DP to the aggregated parameters calculated in each round using the Gaussian mechanism, i.e. adding noise from the Gaussian distribution with mean zero ($\mu=0$) and variance ($\sigma^{2}$) as given in Equation~\ref{eq:dp_fl}. 

\begin{equation}\label{eq:dp_fl}
    \mathcal{N}(\mu, \sigma^{2})=\mathcal{N}\left(0,\left(\frac{n_{e}\cdot C}{m}\right)^{2}\right).
\end{equation}

In this sense, by adding DP we reduce the risk of membership inference attacks (by the definition itself) and also prevents reconstruction attacks on the training data from the aggregated weights. 

Concerning adding personalized DP in FL schemes, in \cite{9449232} the authors proposed a FL scheme in which each client can send their local updates under differential privacy with personally selected privacy levels. In \cite{shen2023pldp}, the authors propose a methodology in which each client defines its privacy parameters and define an unified perturbation algorithm. However, in our approach we aim to customized the privacy budget taking into account a re-identification risk calculated for each client and normalized in the server side. Also we aim to compare this approach with the global DP one mentioned above.

\section{Privacy preserving client drift detection}\label{sec:drift}

The collaboration derived from the application of a federated learning architecture requires a clear consensus on the data used to train the models. Suppose a medical imaging use case, if different devices are used to capture the images, there may be a drift between the data of some clients and others, because they present significant differences to enter a global model with the participation of both. In this case data standards and interoperability present one of the biggest barriers that can be found when training deep learning models \cite{razzak2018deep}. 

In this line, poisoning attacks may arise from two different perspectives: intentionally or unintentionally. In the first case, it is an attacker acting as a client sending maliciously modified model updates to damage the overall performance of the model during aggregation. In other cases, it may be due to the client using training data that differs from that expected or used by the other clients. Returning to the case of medical imaging, this can happen in the collaboration of different institutions taking imaging data with different equipment, or suffering from wear and tear of the equipment itself over time.

The idea of concept drift was already mentioned in \cite{SHI20221168}, but exploring different  optimization strategies for mitigating its impact. However, in this section we aim to provide a formal definition and a methodology to detect it while maintaining data privacy.

However, detecting this issue requires direct access to the data in order to detect anomalies in the client data, which would breach the principles of privacy preserving machine learning that we seek to establish with this architecture. In this sense, in this section we propose a solution to perform client drift detection in a privacy preserving way, without accessing client data directly. We propose a definition of client drift as follows:

\begin{definition}
Be $N$ the number of clients of a FL architecture. Let $d_{i}^{(r)}$ denote the distribution of the data of the client $i$ ($\forall i \in \{1,\hdots,N\}$) in round $r$. Be $P(d_{i}^{(r)})$ the probability distribution. We define client drift as a significant shift in the set $\{P(d_{i}^{(r)}):\forall  i \in \{1,\hdots,N\}\}$. Then, we can consider $P(\overline{d}_{i}^{(r)})$ as the joint (marginal) distribution of all the clients except client $i$ in the system in round $r$, with $n_{k}$ the number of data of client $k$ as follows:

$$
P(\overline{d}_{i}^{(r)})=\sum_{k=1, k\neq i}^{N}\frac{n_{k}}{\sum_{j=1, j\neq i}^{N} n_{j}} P(d_{k}^{(r)}).
$$

Then, client $i$ presents client drift if $P(d_{i}^{(r)}) \not\sim P(\overline{d}_{i}^{(r)})$.

To quantify whether or not these distributions show a significant divergence, we can use proxy heuristics based on classical metrics such as the Kullback-Leibler or Jensen-Shannon divergence, or hypothesis tests on them.
\end{definition}

In this sense, in this study we propose the possibility that, once the weights are received after training the initial model locally on all the clients in each round $r$, the server performs a divergence analysis of the models using the weights that define them \cite{zhang2021client}. To measure the distance or divergence between clients, different distance metrics can be used, such as cosine similarity, Chebyshev distance, Minkowski distance, or Hamming distance among others \cite{chomboon2015empirical}. However, focusing on the differential fact of working directly with the model weights, we propose the use of the following metric to quantify the divergence between pair of models (tested to assess the divergence between clients in \cite{diaz2024personalized}). Be $||\cdot||_{F}$ the Frobenius norm, and $w_{i}^{(r)}$ the weights obtained when training with the data from client $i$ ($\forall i \in \{1,\hdots,N\}$ with $N$ the number of clients) in round $r$, we define the divergence in a symmetric way $d_{i,j}^{(r)}$ as follows for the clients $i$ and $j$:

\begin{equation}
	d_{i,j}^{(r)}=\frac{||w_{i}^{(r)}-w_{j}^{(r)}||_{F}}{\frac{1}{2}(||w_{i}^{(r)}||_{F}+||w_{j}^{(r)}||_{F})}.
	\label{eq:divergence}
\end{equation}

In some cases, it would be enough to perform the client drift detection every certain number of rounds instead of in each round (this frequency should be set in advance depending on the regularity with which new data are expected or the required rate of testing for intentional attacks).

With this method we seek to prevent poisoning attacks by detecting anomalies in the weights sent by each client. These attacks can occur intentionally by malicious attackers seeking to degrade the overall performance of the model, or unintentionally by clients that drift due to labeling errors or changes in the distribution of their data. By detecting whether a client presents drift, we can eliminate it from the training process in the round in which it is detected. Similarly, we can choose to perform clusters of clients based on their similarity. In particular, with the proposed approach we allow detecting these attacks using the concept of client drift while maintaining the privacy principles of an FL architecture, i.e. without accessing the raw data of each client.

\section{Methodology}\label{sec:methodology}

The methodology proposed for the effective development of a privacy preserving machine learning architecture involving different data owners (and therefore applying FL), each of them using sensitive tabular data, is defined as follows:

\paragraph{Anonymization}
(1) First, each client's data is anonymized locally, applying common hierarchies.
(2) The transformation applied to each client is then extracted and harmonized to take a global transformation that satisfies the conditions required for each client. This can be done using \textit{anjana}. The number of suppressed records in each must be taken into account when carrying out the harmonization process.

\paragraph{Client drift detection}
(3) Regarding the detection of poisoning attacks, two scenarios are proposed depending on the type of data. If the data does not change over time, this analysis can be carried out with a first training of the model locally on each client. Thus, in this first inspection, malicious clients that do not behave as expected can be eliminated. For this purpose, the metric proposed in Equation~\ref{eq:divergence} can be calculated for each pair of clients. Then a threshold can be set to quantify if there is a substantial deviation taking into account the divergence between different participating clients.

In case the clients are updating their data in successive batches, it will be necessary to calculate this metric in different rounds, (setting a ratio for its calculation every $x$ rounds), in order to ensure that there are no deviations derived from intentional or unintentional poisonings attacks. In any case, this check can be done with a certain frequency also in cases of data updated in real time.

\paragraph{Federated learning with differential privacy}
(4) To apply DP in the model weights in FL architecture we need to calculate the clipping norm for such parameters. To do this, we will train the model in each client, aggregate it without DP, and run the training again from the aggregated model. Thus, we can calculate the difference between the new local model and the aggregated one. We will take the 90th percentile in order to estimate the clipping norm that will be used for adding the Gaussian noise.
(5) Next, assuming we have a trusted server, we move on to federated training. 
(6) For both performing the FL training and applying DP with fixed clipping in the server side (GDP), we can use \textit{flower}. For this, we have to introduce the noise multiplier, the clipping norm calculated above and the number of sampled clients. Once the models are aggregated, in the classic global DP (GDP) approach, Gaussian noise will be added following the distribution given in Equation~\ref{eq:dp_fl}. 

In this paper, in order to add more privacy to the weights of the models of clients with a higher risk of re-identification, we propose the selection of adaptive and personalized privacy budgets based on a re-identification metric. Specifically, we consider as metric the average re-identification risk of an individual in an equivalence class. We define this metric (in the following noted as ARIREC) as follows:

\begin{definition}\label{def:arirec}
    Be $N$ the number of equivalence classes in a dataset given a set of quasi-identifiers. For each equivalence class $i\in \{1,\hdots,N\}$, let $n_{i}$ be the number of records in the equivalence class. Taking into account the definition of k-anonymity and equivalence class, the risk of identifying an individual in an equivalence class $i$ is given by $1/n_{i}$. Then, we define the average re-identification risk of the equivalence classes (noted as $\mathcal{A}$) as follows:
\end{definition}
\begin{equation}
    \mathcal{A}(N,(n_{1},\hdots, n_{N})) = \frac{1}{N}\sum_{i=1}^{N}\frac{1}{n_{i}}.
\end{equation}

Be $\epsilon$ the privacy budget fixed for the GDP approach. Be $m$ the number of clients and $\mathcal{A}_{i}$ the value of the ARIREC metric introduced in Definition~\ref{def:arirec} for client $i \in \{1,\hdots,m\}$. Associated to each client $i$, in the personalized DP approach we will add noise based on a personalized privacy budget for each client $i$, $\epsilon_{i}$. However, in the practical set up, instead of introducing the value of $\epsilon$, we pass a value for the noise multiplier (noted as $n_{\epsilon}$, see Equation~\ref{eq:dp_fl}). Then, let us consider $\widetilde{\mathcal{A}}_{i}=\frac{\mathcal{A}_{i}}{\sum_{i}^{m}\mathcal{A}_{i}}$. The value of the noise multiplier associated with client $i$ ($n_{\epsilon, i}$), will be calculated as $n_{\epsilon, i}=m\cdot n_{\epsilon}\cdot \widetilde{\mathcal{A}}_{i}$, with $n_{\epsilon}$ the noise multiplier fixed initially.

Thus, the clients will transmit to the server their ARIREC metrics (given by $\mathcal{A}_{i}$ for each client $i \in \{1,\hdots,m\}$), so that the server can normalize it and add the corresponding noise according to the value of the initial noise multiplier. Then, the server will add this noise to the weights received (remember that we assume that the server is trusted but not all clients), and aggregate these weights. Once this is done, the FL process continues as in the classical scheme presented in Section~\ref{sec:fl}.

\section{Experimental Setup}\label{sec:experimental}

In order to test the feasibility of the methodology proposed, we have tested it within an example concerning the use of medical records for cancer severity prediction. 

\subsection{Data}

The data used within this experimental setup is openly available in \cite{data}, and it contains information created synthetically concerning global cancer patient data from 2015 to 2024. 

Specifically, 50000 records were available, including variables associated with patient demographic parameters (\textit{age}, \textit{gender} and \textit{country}), risk-related variables (\textit{genetic risk}, \textit{air pollution}, \textit{alcohol use}, \textit{smoking} and \textit{obesity level}) and clinical ones (\textit{cancer type}, \textit{cancer stage}, \textit{survival years} and \textit{severity score}) among others.

The objective in this experiment is to predict the target severity score of each patient. To do so, we will use the aforementioned risk-related variables, the type of cancer and stage, the year and patient-related variables such as age and gender (we will see that after anonymizing the data this variable is eventually suppressed in our setup). 

In order to implement a federated learning architecture, we have divided the initial dataset into several disjoint clients according to the country of origin of the patients represented in the database (10 clients). This aims to simulate a use case involving the collaboration of hospitals from different countries grouped by country. The distribution of the data across the different countries is shown in Table~\ref{tab:clients_summary}.

\subsection{Model}

For this example we have created a simple neural network based regression model to predict the target severity score of each patient. The model has 4 dense layers, the first three with \textit{relu} activation (the first one receives the 8 input features) and with 64, 128 and 64 neurons respectively. Finally, the output dense layer has a single neuron. The model uses the mean square error as loss and the Adam optimizer with a learning rate of 0.001. Concerning model training and testing, the hyper-parameters fine tuning was performed in the client with more data using a part of the train data as validation. Finally, in order to train the model on each client 80$\%$ of the data were used and the remaining 20$\%$ were used for testing. In all the proposed scenarios, the model has been trained for 100 epochs.

Detailed discussion concerning the model implementation, selection and hyper-parameters fine tuning is intentionally omitted, as optimizing the model performance is beyond the scope of this study.

\subsection{Client drift detection}

After splitting the initial data on artificial clients, it is essential to check whether drift is occurring between clients as described in Section~\ref{sec:drift}. In this case, as the data is static (no new data will be added during the FL rounds), it is sufficient to perform this check at the beginning of the process. To do this, prior to the first round, a client drift test can be performed before starting the federated training, and if detected, the client will be removed from the architecture.

\begin{figure}[ht]
    \centering
    \includegraphics[width=\linewidth]{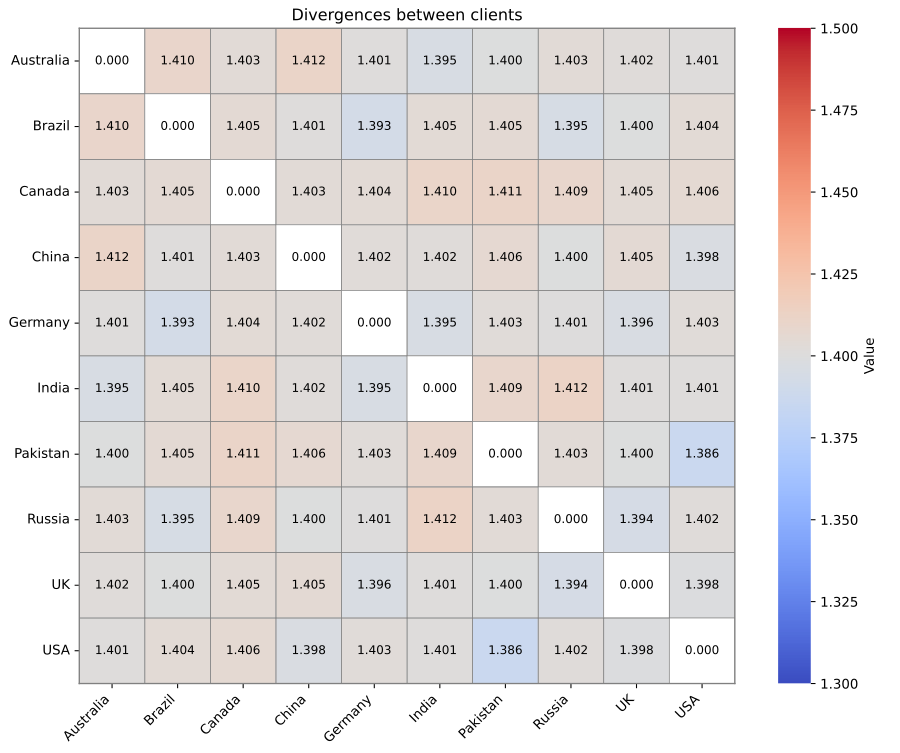}
    \caption{Divergence between clients calculated using Equation~\ref{eq:divergence}.}
    \label{fig:divergences}
\end{figure}

Thus, a first inspection is carried out by training the model for 100 epochs (corresponding to one round) and it is obtained that, after comparing the different pairs of clients with respect to the metric presented in the Equation~\ref{eq:divergence}, none presents significant differences with respect to the rest, so that no client drift is detected. This can be seen in Figure~\ref{fig:divergences}. 

\subsection{Data anonymization}

Then we proceed with the data anonymization process, selecting as identifier the \textit{`Patient ID'} column available in the original database and as quasi-identifiers (QIs) \textit{`Age'}, \textit{`Gender'}, \textit{`Year'} and \textit{`Cancer Type'}, since they are the variables that an attacker could a priori know about an individual represented in the database. Regarding the hierarchies applied, as a first level we allow to generalize the age at intervals of 5 years, and at intervals of 10 years as second level. In addition, we allow to suppress the gender if needed. On the other two QIs we do not apply any hierarchy since we are interested in keeping them unchanged for training the ML model. In addition, we set a suppression limit of 10$\%$ and apply \textit{k-anonymity} with $k=5$.

Once we have anonymized the data, we obtain the transformations. If necessary we would harmonize them, however, in all cases we have obtained the same transformation: [2,1,0,0], i.e., the gender variable is suppressed and age is generalized to 10-year intervals.

Now we proceed to check the level of anonymity with respect to other metrics, considering the variable to be predicted as sensitive attribute. Specifically, in Table~\ref{tab:clients_summary} we show the value of $t$ obtained for \textit{t-closeness}, which allows us to intuitively know the difference between the distribution of the sensitive attribute (the label) in each database for each country and with respect to each equivalence class where the QI are identical. The lower the value of $t$, the greater the privacy, but we may also lose accuracy when developing a model. Specifically, we can note that these values are between 0.20 and 0.32 for all the clients analyzed. 

Finally, we calculate in each case the average re-identification risk by equivalence class (ARIREC or $\mathcal{A}_{i}$ for each client $i \in \{1,\hdots,10\}$). This value will be used to determine the personalized privacy budget associated with each client. The value of the ARIREC metric for each client is shown in Table~\ref{tab:clients_summary}.

\begin{table}[ht]
    \centering
    \caption{Clients summary (clustered by countries): number of data, ARIREC and value of $t$ for $t$-closeness.}
    \label{tab:clients_summary}
     \begin{tabular}{cccc}
     \toprule
         \textbf{\textit{Client}} & \textit{$\#$Data} & \textit{ARIREC} & \textit{t-closeness} \\
         \midrule
            \textit{Australia} & 5092 & 0.1168 & 0.2102 \\
            \textit{Brazil} & 5004 & 0.1180 & 0.2027 \\
            \textit{Canada} & 4864 & 0.1197 & 0.2426 \\
            \textit{China} & 4913 & 0.1214 & 0.2683 \\
            \textit{Germany} & 5024 & 0.1183 & 0.2918 \\
            \textit{India} & 5040 & 0.1179 & 0.2420 \\
            \textit{Pakistan} & 4926 & 0.1200 & 0.3113 \\
            \textit{Russia} & 5017 & 0.1203 & 0.2713 \\
            \textit{UK} & 5060 & 0.1173 & 0.2558 \\
            \textit{USA} & 5060 & 0.1190 & 0.2363 \\
    \bottomrule
    \end{tabular}
\end{table}

\subsection{Personalized privacy budget}
Concerning personalized DP, first of all we must set an initial privacy budget. Specifically, following the Equation~\ref{eq:dp_fl}, we will set a noise multiplier ($n_{\epsilon}$), which is a value inversely related to $\epsilon$. In addition, we must set a clipping norm ($C$). For the value of the noise multiplier different values have been tested in order to obtain a reasonable balance in the noise added in this example. Finally, a value of 0.05 was selected for the case of global DP (GDP) applied directly on the aggregated weights (this will be further explained in the next section). For the clipping norm, the methodology proposed in Section~\ref{sec:methodology} has been followed and a value of 8.5 has been taken, as it is slightly higher than the value calculated for percentile 90 of the difference between the local weights of round two and those added after the first round. 

\subsection{Model training: results and evaluation}

In order to test the methodology in this experimental setup, three settings are tested: standard (vanilla) FL training, FL with GDP in the aggregation process and FL with personalized DP (PDP) with customized privacy budget. In the second case, DP is added to the weights once aggregated, in the third case, in order to add a customized noise multiplier, it is applied to each vector of local weights (once clipped) and then aggregated. The training is performed in all cases for 5 rounds.

Thus, the results of these three studies with respect to the mean absolute error (MAE) and the mean relative error (MRE) are shown in Table~\ref{tab:mre_results}, showing the mean and the standard deviation obtained for these metrics by repeating the experiment 10 runs.

However, it is clear that setting the same value (on average) of $n_{\epsilon}$ and $n_{\epsilon,i}$ in both cases does not add the same amount of noise, but they are different approaches to protect the updates sent by the clients for aggregation and transmission of the aggregated model. However, in order to add a comparable amount of noise in both cases, the following has been taken into account:

Be $w_{i}$ the weights obtained after training locally in each client $i$ $\forall i\in\{1,\hdots,m\}$, $w_{agg}$ the aggregated weights (by simplicity assume that we are using \textit{FedAvg}) without DP. Be $w_{i}^{(PDP)}$ the weights of client $i$ after adding personalized DP, then $w_{i}^{(PDP)}=w_{i}+\mathcal{N}\left(0,\left(\frac{n_{\epsilon}\cdot C \cdot \widetilde{\mathcal{A}}_{i} \cdot m}{m}\right)^{2}\right)$ $\forall i \in \{1,\hdots, m\}$. Be $w_{agg}^{(PDP)}$ the aggregated weights after adding personalized DP (PDP) in the weights of each client $i$, and be $m_{i}$ the number of data of client $i$: $w_{agg}^{(PDP)} = \sum_{i=1}^{m}\frac{m_{i}}{\sum_{i=1}^{m}m_{i}}w_{i}^{(PDP)}$. For simplicity, let us assume that $\widetilde{\mathcal{A}}_{i} \cdot m = 1$ and $m_{i}=m_{j}$ $\forall i,j \in \{1,\hdots, m\}$, $i\neq j$. Also, be $w_{agg}^{(GDP)}$ the weights aggregated adding global DP once aggregated (classic approach). Then: 

\begin{align*}
     w_{agg}^{(GDP)} & =w_{agg}+\mathcal{N}\left(0,\left(\frac{n_{\epsilon}\cdot C }{m}\right)^{2}\right) \\
     w_{agg}^{(PDP)} & = \frac{1}{m}\sum_{i=1}^{m}\left(w_{i}+\mathcal{N}\left(0,\left(\frac{n_{\epsilon}\cdot C }{m}\right)^{2}\right)\right) =\\
    & = w_{agg}+\mathcal{N}\left(0,\left(\frac{1}{\sqrt{m}}\frac{n_{\epsilon}\cdot C }{m}\right)^{2}\right).\\
\end{align*}

Then, in order to compare the noise added in the cases of $w_{agg}^{(PDP)}$ and $w_{agg}^{(GDP)}$, we will fix a value for the $n_{\epsilon}$ value for $w_{agg}^{(GDP)}$, and in the case of $w_{agg}^{(PDP)}$ we will multiply such value by $\sqrt{m}$ (and also by $\widetilde{\mathcal{A}}_{i} \cdot m$, since in the previous calculations we assume this value to be one only for simplicity of the analysis).

\begin{table}[ht]
    \centering
    \caption{Comparison of the MAE and MRE obtained in the last round of the training (10 runs): vanilla FL, FL with DP aggregation and FL with personalized privacy budgets for DP aggregation.}
    \label{tab:mre_results}
    \begin{tabular}{lcc}
        \toprule
         \textbf{Training scenario} & \textit{MAE (mean $\pm$ std)} & \textit{MRE (mean $\pm$ std)}  \\
         \midrule
         \textit{\textbf{FL (vanilla)}} & 0.578 $\pm$ 0.006 & 0.125 $\pm$ 0.002 \\ 
         \textit{\textbf{FL+Global DP}} & 0.847 $\pm$ 0.273 & 0.182 $\pm$ 0.057 \\ 
         \textit{\textbf{FL+Personalized DP}} & 0.836 $\pm$ 0.270 & 0.177 $\pm$ 0.051 \\ 
         \bottomrule
    \end{tabular}
\end{table}

In view of Table~\ref{tab:mre_results}, we can observe how using the personalized privacy budgets on each individual vector of weights instead of directly on the aggregated ones, has a lower impact on the performance of the resulting FL model both in terms of MAE and MRE, as well as providing more or less noise to each client contribution according to the established risk metric.

Finally, it is important to note that we control the privacy level through the noise multiplier. Specifically, $\epsilon$ can be estimated a posteriori using standard DP accounting, but our focus in this preliminary study is on relative privacy allocation across clients, and this task is proposed as future work applied to real scenarios with stricter privacy levels.

\section{Conclusions and future work}\label{sec:future}

The objective of this paper is to present a complete methodology to train a ML/DL model on tabular data from multiple decentralized clients using a complete FL architecture with additional privacy measures. Specifically, we intend to prevent the most common types of attacks on these architectures using anonymization, differentially private aggregation methods assuming trusted server, and client drift detection. Additionally, our methodology proposes to use personalized privacy budgets to add DP on local weights based on a metric that quantifies the risk of re-identification of different clients.

To illustrate the performance of the proposed methodology, an experiment based on tabular data related to medical records has been proposed. We have analyzed the results obtained in three setups, including vanilla FL, FL with GDP and FL with differential privacy with personalized noise multipliers based on the proposed ARIREC metric. By using this methodology we take into account the risk of each client to add more or less noise to their model updates. In the use case analyzed we can note that the results obtained with the proposed approach are slightly better than those obtained with the classical global DP one. 

As for future work, and taking into account the limitations of this preliminary study, the first step involves extending this workflow to other use cases, even involving non-i.i.d clients. Secondly, we will extend the methodology to cases where the server is not trusted. In this line, we are interested in exploring the integration of homomorphic encryption for the secure transmission of weights and their encrypted aggregation. Specifically, multi-key homomorphic encryption emerges as a promising solution for this case, allowing different clients to use their own public and private keys to do the aggregation, assuming a semi-honest server. Finally, concerning the use of DP, an extended work will focus on providing $\epsilon$ and $\delta$ parameters accounting and composition analysis throughout the FL rounds. 

\vspace{-0.2cm}

\section*{Acknowledgment}

The authors would like to thank the funding through the EOSC SIESTA project ``Secure Interactive Environments for Sensitive daTa Analytics'' that has received funding from the European Union's Horizon Europe research and innovation programme under grant agreement number 101131957.

\vspace{-0.1cm}
\bibliographystyle{IEEEtran} 
\bibliography{bib}

\end{document}